\documentclass[10pt,conference,letterpaper]{IEEEtran}
\usepackage{times,amsmath,epsfig}
\usepackage[]{algorithmic, algorithm}
\title{On Breaching Enterprise Data Privacy Through Adversarial Information Fusion}
\author{%
{Srivatsava Ranjit Ganta, Raj Acharya}%
\vspace{1.6mm}\\
\fontsize{10}{10}\selectfont\itshape
Dept of Computer Science, Penn State University\\
University Park, PA, USA\\
\fontsize{9}{9}\selectfont\ttfamily\upshape
ranjit@cse.psu.edu\\
acharya@cse.psu.edu%
\vspace{1.2mm}\\
\fontsize{10}{10}\selectfont\rmfamily\itshape
\fontsize{9}{9}\selectfont\ttfamily\upshape
}
\begin{document}
\maketitle
\begin{abstract} 
Data privacy is one of the key challenges faced by enterprises today. \emph{Anonymization} techniques address this problem by \emph{sanitizing} sensitive data such that individual privacy is preserved while allowing enterprises to maintain and share sensitive data. However, existing work on this problem make inherent assumptions about the data that are impractical in day-to-day enterprise data management scenarios. Further, application of existing anonymization schemes on enterprise data could lead to adversarial attacks in which an intruder could use \emph{information fusion} techniques to inflict a privacy breach. In this paper, we shed light on the shortcomings of current anonymization schemes in the context of enterprise data. We define and experimentally demonstrate \emph{Web-based Information-Fusion Attack} on anonymized enterprise data. We formulate the problem of \emph{Fusion Resilient Enterprise Data Anonymization} and propose a prototype solution to address this problem.  
\end{abstract}

%
\section{Introduction}
Data privacy is one of the key challenges faced by enterprises today. Sensitive individual-specific information such as customer data, employee data etc are maintained and used for various purposes. Several instances of data privacy breaches~\cite{agrawal-hippocratic} in the recent past have resulted in financial as well as reputation losses for enterprises. \emph{Anonymization} techniques address this problem by \emph{sanitizing} sensitive data such that individual privacy is preserved while allowing enterprises to maintain and share sensitive data. Recently, there has been a lot of work~\cite{SamaratiandSweeney1998}~\cite{LeFevreDR06}~\cite{Machanavajjhala2006}~\cite{agrawal-ppdm}~\cite{evfimievski02} on data anonymization schemes. These techniques can be broadly classified into two types:
\begin{itemize}
	\item \textbf{\emph{Partitioning} based anonymization schemes :} The first class of techniques guarantee privacy by partitioning the data such that an adversary cannot uniquely identify the individuals falling in each partition. The basic ideology behind these techniques is \emph{blending in the crowd} which guarantees that an individual or entity cannot be distinguished from a minimum number of other people. $K$-anonymity~\cite{SamaratiandSweeney1998}, $l$-diversity~\cite{Machanavajjhala2006} and other work in this line~\cite{LiLV07} achieve partitioning through \emph{generalization} and \emph{suppression} techniques. On the other hand, techniques such as~\cite{AggarwalFKKPTZ06},~\cite{Ferrer2002} achieve this by \emph{clustering} the data. Partitioning based solutions are mainly applied to \emph{non-interactive} scenarios where the data needs to be \emph{published/released} after anonymization.  
	\item \textbf{\emph{Perturbation} based anonymization schemes :} The other class of techniques guarantee data privacy by adding \emph{noise} to the sensitive data and thus preventing identification. Solutions in this category can be further classified based on whether the setting considered is interactive or not. Solutions such as~\cite{agrawal-ppdm}~\cite{evfimievski02} add noise to perform specific data mining tasks in a non-interactive setting. More recent solutions such as~\cite{Dwork06} add randomized noise in an \emph{interactive} setting where-in the particular function to be evaluated on the data is known \emph{apriori}.
\end{itemize}

In this paper, we consider a \emph{non-interactive} setting where the data needs to be \emph{released}/\emph{published}. We focus on partitioning based schemes as they are readily applicable to generic databases including data with categorical attributes. Table~\ref{sensitivedata} depicts a typical individual-specific data considered in partitioning based anonymization literature. Observe that there exists a \emph{classification} of data attributes (as shown in Table~\ref{sensitivedata}) into three different types: 
\begin{enumerate}
\item \textbf{Identifier Attributes:} Attributes carrying explicit identifiers such as \emph{Name}, \emph{SSN} etc. 
\item \textbf{Quasi Identifier Attributes:} Attributes that could indirectly lead to identification of individuals in the database such as \emph{Age}, \emph{Zipcode} and \emph{Gender} etc. These are also sometimes referred to as \emph{Non-Sensitive} attributes.
\item \textbf{Sensitive Attributes:} Attributes carrying the sensitive information about the individuals such as \emph{Disease}, \emph{Income} etc. 
\end{enumerate}
Based on this classification, existing solutions assume that the \emph{Identifier Attributes} in the database are stripped \emph{prior} to the anonymization process. This was under the implicit assumption that the identifier attributes were necessary neither for the release nor for the intended purpose of the release. We believe that this assumption is too restrictive and is even impossible in some scenarios where the presence of explicit identifiers is \emph{necessary} for the intended purpose of the anonymized release~\cite{Ganta08SAC}. Consider the following scenario:
\begin{table*}[!htb]
\centering
	\begin{tabular}{||c|c||c|c|c||c||} \hline
		  \multicolumn{2}{||c||}{\textbf{Identifiers}} & \multicolumn{3}{c||}{\textbf{Quasi Identifiers}} & \textbf{Sensitive} \\ \hline
			\textbf{Name} & \textbf{SSN} &\textbf {Zipcode}	&\textbf {Age}	&\textbf {Nationality} &\textbf {Condition} \\ \hline
			Alice & 111-111-1111	&13053 &28	&Russian &AIDS\\ \hline
			Bob & 222-222-2222 &13068 &29  &American &Flu\\ \hline
			Christine & 333-333-3333	&13068 &21	&Japanese	&Cancer\\ \hline
			Robert & 444-444-4444 &13053 &23	&American	&Meningitis\\ \hline
	\end{tabular}
\caption{Sensitive Database} 
\label{sensitivedata}
\end{table*}

\textbf{Enterprise Data - Example :} Table~\ref{enterprisedata} depicts a customer database in a typical financial institution. The data contains names of all the customers along with certain \emph{non-sensitive} and \emph{sensitive} information. The non-sensitive attributes are: \emph{Investment Volume Index (Invst Vol)} to indicate the volume of investment (number of shares traded etc.) made by the customer in the past, \emph{Investment Amount Index (Invst Amt)} to indicate the amount of investment (amount involved in previous trades etc.) made by the customer in the past, \emph{Customer Valuation (Valuation)} to indicate the assigned value of the customer. The only \emph{sensitive} attribute, \emph{Customer Personal Income (Income)}, corresponds to the customer's personal income. Databases such as this are an integral part of enterprises and are maintained and used for key operations everyday. In this paper, we shall refer to them as \emph{Enterprise Databases}.

The internal \emph{release} of such data along with explicit identifiers (\emph{Customer Names}) is a necessity for several enterprise operations such as accounting, record keeping etc. However, at the same time, such a release should not compromise the privacy of sensitive information (\emph{Customer Personal Income}). Note that trivial solutions such as removal of identifiers or use of pseudonyms are not viable in such scenarios. The key properties here are:
\begin{itemize}
\item The inclusion of identifying information is necessary for the \emph{release} to serve the intended purpose.
\item Sensitive data disclosure should not be compromised even in the presence of explicit identifiers.
\end{itemize}

\begin{table}[h]
	\centering
		\begin{tabular}{||c|c|c|c|c||} \hline
			\textbf{Name}	&\textbf {Invst Vol}	&\textbf {Invst Amt}	&\textbf {Valuation} &\textbf {Income} \\ \hline
			Alice	&8	  &7	&4 &$$91,250\\ \hline
			Bob	&5	   &4	&4 &$$74, 340\\ \hline
			Christine	&4	&5	&5 &$$75,123\\ \hline
			Robert	&9	&8	&9 &$$98,230\\ \hline
		\end{tabular}\caption{Enterprise Data}
	\label{enterprisedata}		
\end{table} 
In the enterprise database scenario described above, anonymizing data using existing techniques falls short in providing adequate protection against adversarial attacks. This is because existing techniques ~\cite{SamaratiandSweeney1998}~\cite{LeFevreDR06}~\cite{Machanavajjhala2006} make an assumption that \emph{Identifier Attributes} are stripped \emph{prior} to the anonymization process. Consider the possibility in which an adversary (possibly an \emph{insider}) is given (or otherwise acquires) access to the anonymized release of an enterprise database. Now, the adversary can use the identifiers present in the release to collect \emph{auxiliary} information about the individuals present in the database from a multitude of sources such as the web (homepages, blogs etc). The adversary could then \emph{fuse} the auxiliary information with the anonymized release to estimate sensitive data. 
 
\textbf{Web-Based Information-Fusion Attack :} Consider the enterprise data example described earlier as shown in Table~\ref{enterprisedata}. One way to internally release this table is to remove the customer salary information and publish the non-sensitive data as it is. The problem with this approach is that one can estimate the sensitive data based on the non-sensitive information present in the release. The solution is to anonymize the non-sensitive information and remove the sensitive information. Table~\ref{existingresult} shows the anonymized release of this data using partitioning based anonymization scheme such as $K$-anonymity proposed by Sweeney et al.~\cite{SamaratiandSweeney1998}. We use $K$-anonymization as a representative of partitioning based solutions for data anonymization as other solutions in this category produce similar results.  
\begin{table}[h]
	\centering
		\begin{tabular}{||c|c|c|c|c||} \hline
			\textbf{Name}	&\textbf {Invst Vol}	&\textbf {Invst Amt}	&\textbf {Valuation} &\textbf {Income} \\ \hline
			Alice	&[5-10]	   &[5-10]	&[1-5] &- \\ \hline
			Bob	&[5-10]   &[1-5]	&[1-5] &- \\ \hline
			Christine	&[1-5]	&[1-5] 	&[1-5]   &- \\ \hline
			Robert	&[5-10]	&[5-10]	&[5-10] &- \\ \hline
		\end{tabular}	\caption{Anonymized Enterprise Data}
	\label{existingresult}
\end{table}
\begin{table}
	\centering
		\begin{tabular}{||c|c|c||} \hline
			\textbf{Name}	&\textbf {Employment}	&\textbf {Property Holdings}	\\ \hline
			Alice	&CEO, Deutsche Bank 	  &3560	\\ \hline
			Bob	&Manager, Verizon  &1200	\\ \hline
			Christine	&Assistant, NYU	&720	\\ \hline
			Robert	&CEO, Microsoft	&5430	\\ \hline
		\end{tabular}\caption{Auxiliary Data Collected By The Adversary}
	\label{webdata}		
\end{table}

Table~\ref{existingresult} is now deemed \emph{safe} and is released internally within the enterprise. Now, consider the scenario in which an adversary employee \emph{Bob} is granted access to this anonymized release. Note that the release does not give \emph{Bob} the sensitive information i.e customer personal income data. However, he has access to non-sensitive information such as the customer valuation, investment volume etc. \emph{Bob}'s goal is to use the anonymized release to estimate the customer personal income values. To achieve this, he uses the customer names present in the release to search for additional information about the customers available on the web which will help him estimate their personal income. For example, he collects information about the customer's \emph{Employment}, \emph{Property Holdings} etc. Example of such data collected from the web is shown in Table~\ref{webdata}. Now, by \emph{fusing} this information with the anonymized release the adversary can estimate the sensitive customer personal income information. In this example, let's say the income range for all the customers is [\$40000 - \$100000] and could be divided into three classes \emph{Low} [\$40000 - \$60000], \emph{Medium} [\$60000 - \$80000], and \emph{High} [\$80000 - \$100000]. Now, consider the customer \emph{Robert}. With an estimated valuation falling in the highest range [5-10], \emph{Bob} concludes that \emph{Robert} falls into the highest income category [\$80000 - \$100000]. By looking at his employment and property holdings (and possibly other auxiliary information), \emph{Bob} can further improve his estimate and conclude that \emph{Robert} falls into upper category [\$90000 - \$100000] of the \emph{High} income class. Based on this, he estimates that \emph{Robert's} salary is the average of range [\$90000 - \$100000] i.e \$95000. This example demonstrates, how, by using the auxiliary information obtained from the web an adversary could obtain a close estimate of \emph{Robert's} actual income. Although in the above example the attacker uses his understanding of the data to \emph{fuse} the anonymized release with web data, in reality, he could use various \emph{Information Fusion} techniques for this purpose. Information Fusion is a well-studied paradigm in which multiple data sources are used to improve knowledge extraction. 

\begin{figure}[!htp]
\centering
\epsfig{figure=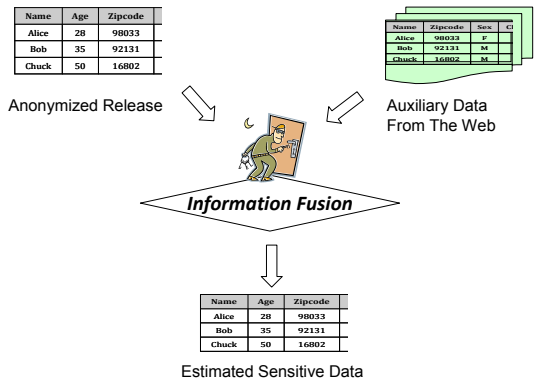, scale=0.4}
\caption{Web-Based Information-Fusion Attack}
\label{attackdiagram}
\end{figure}

In the attack demonstrated above, an adversary with access to anonymized enterprise data gleans auxiliary information from the web and uses information fusion techniques to inflict a privacy breach. In this paper, we refer to such an attack as \emph{Web-Based Information-Fusion Attack} on enterprise data. This is illustrated in Figure~\ref{attackdiagram}. Note that this attack is an example of an attack-model in which a \emph{human-in-the-loop} inflicts a privacy breach. 

\subsection{Contributions and Organization}
In this paper, we demonstrate the shortcomings of existing anonymization schemes when applied to enterprise data through the \emph{Web-Based Information-Fusion Attack}. Our main contribution is the formulation of \emph{Fusion Resilient Enterprise Data Anonymization} problem. We propose an iterative scheme to find an optimal anonymization that offers maximum protection against such attacks for a given dataset.  The rest of the document is organized as follows: Section 2 provides the related work to this problem. Section 3 elaborates on the Web-Based Information-Fusion Attack and discusses the assumptions made regarding the attack. In Section 4, we formulate the problem of Fusion Resilient Enterprise Data Anonymization. We then present our solution strategy to address the problem through incremental anonymization in Section 5. Section 6 presents experimental results by demonstrating the attack on a real data set and presenting the prototype solution. Section 7 provides the conclusion and future work.

\section{Related Work}
Data privacy has received a lot of attention from both computer science and statistical research communities. In statistical literature, studies on data confidentiality~\cite{Duncan1986}~\cite{Duncan1990} propose the use of matrix masks for anonymizing data. In the computer science literature, several recent studies~\cite{SamaratiandSweeney1998}~\cite{LeFevreDR06} have been done in the context of $K$-anonymity. Ferrer~\cite{Ferrer2002} proposed heuristic algorithms for optimal $K$-anonymization on quantitative data. Several problems with $k$-anonymity based partitioning techniques have been studied in~\cite{Machanavajjhala2006}~\cite{LiLV07} and others. In~\cite{Machanavajjhala2006}, Machanavajjhala et al. pointed to the possibility of attacks on $k$-anonymized data because of lack of diversity in the sensitive values corresponding to each partition. Later,in ~\cite{LiLV07}, Li et al. provided an argument that $l$-diversity is neither a sufficient nor a necessary condition to guard against attacks on $k$-anonymized data. They proposed a scheme in which the distribution of sensitive values with-in each partition should not be far from the distribution of sensitive values in the original data. 

One of the primary challenges in data anonymization is to take into consideration the \emph{auxiliary} information (also called \emph{external knowledge}, \emph{background knowledge} or \emph{side information}) that an adversary can glean from other channels. Recent work on partitioning based techniques~\cite{Machanavajjhala2006}~\cite{MartinKMGH07}~\cite{ChenRL07} has attempted to define adversary's \emph{background knowledge} and possible privacy breach based on this. Martin et al~\cite{MartinKMGH07} provide a first formal treatment of adversarial background knowledge. They propose a language for expressing the adversary's knowledge based on conjunctive propositions. More recently, Chen et al.~\cite{ChenRL07} have attempted to fill this gap, by proposing an extension to the same language based framework. However, these models do not consider auxiliary information obtained \emph{using identifying information present in the anonymized release}. 

On the other hand, there has been some work~\cite{WangF06}~\cite{ByunSBL06}~\cite{XiaoT07} on addressing the problem of anonymizing sequential releases. The problem here is to ensure that the current release of a particular data set does not lead to a disclosure with respect to previous releases on the same data set. Orthogonal to these works,in ~\cite{WongFWP07} Wong et al prove that adversary's knowledge of the anonymization algorithm could lead to a privacy breach. In~\cite{aggarwal2006}, Aggarwal et al. pose the problem of adversarial rule mining attack on anonymized data. Our work is critically different from these studies as we consider inferential attribute disclosure based on \emph{Information Fusion} using external information sources.

\section{Web-Based Information-Fusion Attack}

\subsection{Information Fusion}
In this paper, we use \emph{fuzzy inferencing} to build an Information Fusion system. This section provides a brief introduction to fuzzy inferencing and how it can be used by the adversary to fuse the anonymized release with web-based auxiliary information. 

\begin{figure}[!htp]
\centering
\epsfig{figure=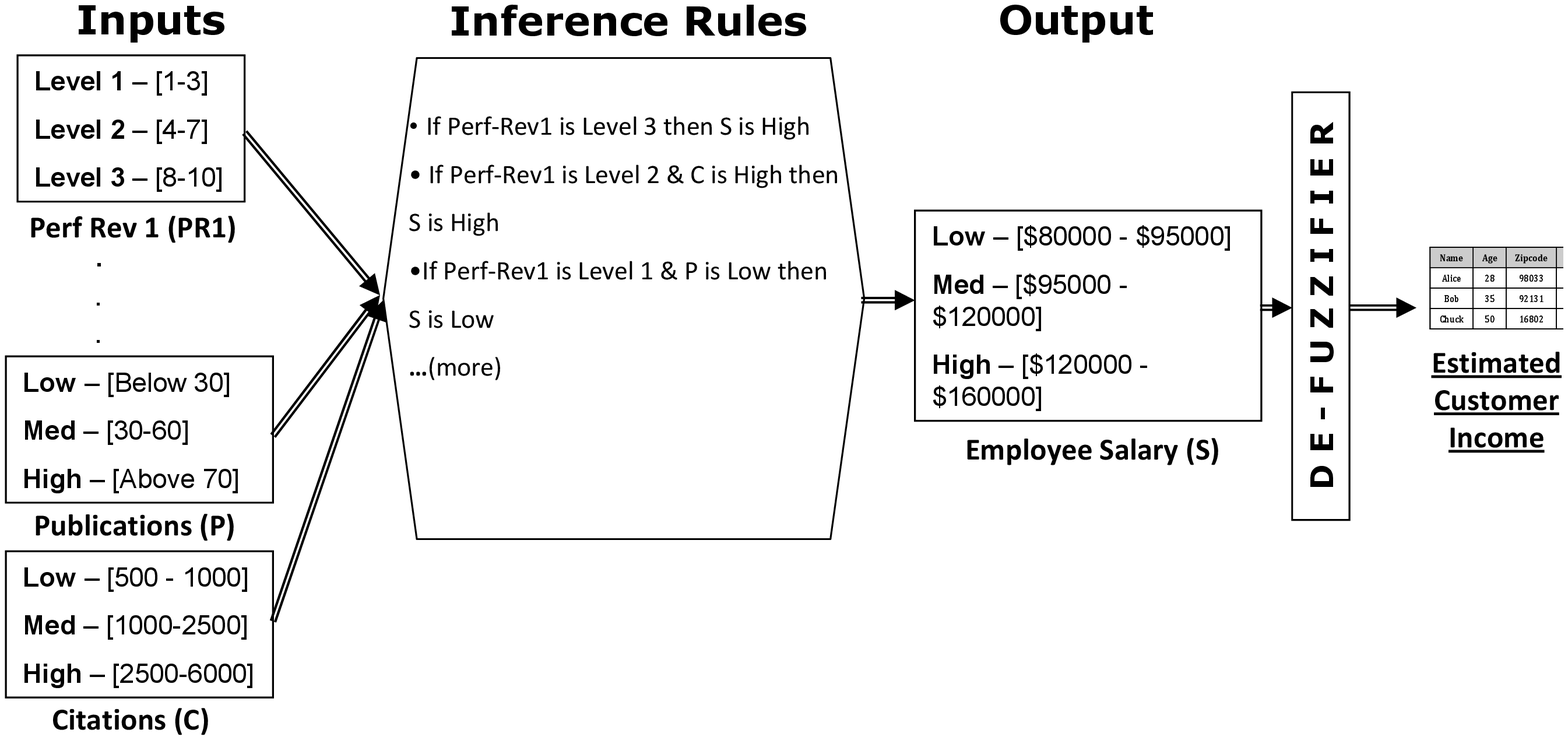, width=250pt, height=125pt}
\caption{Fuzzy Inference System}
\label{newfis}
\end{figure}

\emph{Fuzzy Inference} is a well-studied paradigm based on \emph{fuzzy logic}, \emph{fuzzy if-then rules} and \emph{fuzzy reasoning}. Basically, it provides a mechanism to \emph{map} a set of \emph{inputs} to a set of \emph{outputs} using a set of \emph{rules}. We refer the reader to~\cite{kosko} for an introduction to fuzzy inference systems. The first step involved in creating a fuzzy inference system is to determine the \emph{inputs} and \emph{outputs}. In the web-based information-fusion attack, the inputs include all the data attributes available to the adversary through: 1. The anonymized release and 2. The auxiliary data collected through the web. In our running example from Section 1, the attributes \emph{Investment Volume Index}, \emph{Investment Amount Index}, \emph{Customer Valuation} from the anonymized release in Table ~\ref{existingresult} form the first half of inputs to the information fusion system. The attributes \emph{Employment}, \emph{Property Holdings} collected from the web form the second half of inputs. The output consists of single attribute, \emph{Customer Personal Income}, which the adversary intends to estimate. In the second step, the adversary defines \emph{fuzzy-set definitions} for each of the input and output attributes. He then uses domain knowledge to formulate a set of \emph{rules} mapping the input fuzzy sets to the output fuzzy sets. Figure ~\ref{newfis} illustrates the system.

%
\subsection{Attacker Capability} 

We assume that the intruder is an \emph{insider} who is given or otherwise acquires access to the anonymized data. Thus, the intruder has access to individual identifiers that can be used to index into the web and other data sources. The intruder is assumed to have the domain knowledge about the data to perform information fusion.  

\section{Problem Formulation}
In this section we formulate the problem of \emph{Fusion Resilient Enterprise Data Anonymization} to address web-based information-fusion attacks. Since it is not possible to quantify the amount of auxiliary information the adversary can collect, it is not practical to completely prevent such attacks. However, by estimating the auxiliary information that an adversary could collect, we can  \emph{minimize the extent of privacy breach} in case of such an attack. This forms the primary goal of our problem formulation: For a given sensitive dataset, we need to find an anonymization such that the release causes minimum breach in case of a fusion attack. On the other hand, one of the important factors involved in data anonymization is the \emph{utility} of the release~\cite{Bayardo2005}~\cite{LeFevreDR06}. The utility of an anonymized release is a measure of usefulness of the release for the intended purpose such as a specific task to be performed on the data Ex. Classification etc. Several standard measures such as~\cite{Bayardo2005} have been proposed in the literature to compute data utility.  Hence, the secondary goal of our problem formulation is to maximize data utility. With these goals in hand, we proceed to formulate the overall goal as follows:

Let $P = \{p_{ij}\}_{m \times n}$ be a sensitive private dataset defined over a finite set of attributes $\{P_1, P_2, \ldots, P_n \}$. \\
Let $Q = \{q_{ij}\}_{r\times s}$ be the auxiliary data gathered by the \emph{intruder} from the web over a set of attributes $\{Q_1, Q_2, \ldots, Q_s \}$.\\ 
Now, let $P'$ be a \emph{candidate} anonymization of $P$.\\
Let $F$ be an \emph{information fusion} system that takes in $P'$ and $Q$ as inputs and produces $\hat{P}$, an estimate of $P$.\\
Let $U$ be a measure of utility of $P'$. \\
\textbf{Goal :} The goal of \emph{Fusion Resilient Enterprise Data Anonymization} is to compute \emph{a} $P'$ from $P$ such that:
\begin{enumerate}
	\item $P'$ is \emph{resilient} to \emph{Web-based Information Fusion Attacks}.
	\item The utility $U$ offered by $P'$ meets the release requirements. 
\end{enumerate}

To formulate the problem based on the above goal, we need to quantify the \emph{resilience} to web-based information-fusion attacks. We define this using the following definitions:

\textbf{Definition 1 Dissimilarity ($D_1 \circ D_2$)} \emph{For two datasets $D_1$ and $D_2$ representing the same set of individuals and the same set of attributes, $D_1 \circ D_2$ is a measure of \emph{dissimilarity} between them}. 

For two datasets $\{D_1\}_{m \times n}$ and $\{D_2\}_{m \times n}$ representing the same set of individuals, we compute the dissimilarity using mean square distance $D_1$ and $D_2$: 
\[D_1 \circ D_2 = \frac{1}{m}*Tr((D_1-D_2)^T(D_1-D_2)) \] 	
where $m$ is the total number of records in each database and $Tr(A)$	of a matrix $A$ is the trace of $A$, i.e the sum of the elements of the main diagonal.

As defined earlier, $\hat{P}$ is an estimate of $P$ made by the adversary based on a candidate release $P'$ and web-based auxiliary data $Q$ using the information fusion system $F$. 

\[ \hat{P} = F(P', Q) \]

In order for privacy of $P$ to be protected, the dissimilarity between  $P$ and the estimate made by the adversary, $\hat{P}$, needs to be \emph{large}. The more the dissimilarity $P \circ \hat{P}$, the better protected $P$ is. Also, the dissimilarity between $P$ and $\hat{P}$ \emph{quantifies} the \emph{protection} offered by the corresponding $P'$ against information fusion attacks. Based on this, we now define a \emph{Fusion Resilient Anonymization} as:

\textbf{Definition 2 Fusion Resilient Anonymization} \emph{An anonymization $P'$ of a given sensitive data $P$ is resilient to fusion attacks if the dissimilarity $(P \circ \hat{P})$ between $\hat{P}$ and $P$ is above a certain threshold value $T_p$}. 

So, for a candidate anonymization $P'$ to be a \emph{safe} release, the corresponding $(P \circ \hat{P})$ needs to be above a certain threshold value $T_p$. It is obvious to note that, among all the possible anonymizations ($P'$s) that satisfy this property, the one that has maximum value of $(P \circ \hat{P})$ offers maximum protection. So, for the anonymization $P'$ to offer maximum resilience to web-based information fusion attacks, the \emph{dissimilarity} $(P \circ \hat{P})$ needs to be maximized. 

Recall that in addition to maximizing the protection against information-fusion attacks, the utility of the release ($U$), should be maximized. Let $W_1$ and $W_2$ be the weights assigned by the publisher for privacy protection against information fusion attacks and data utility respectively. Now, the final objective can be stated as a \emph{weighted sum of protection and utility} of the form:
\[W_1*(P \circ \hat{P})+ W_2*U\]

Now, the problem can be stated as,

\textbf{Problem :} Given a private dataset $P$, web-based data $Q$ and an information-fusion system $F$, find the fusion resilient anonymization $P'$ that maximizes $H = W_1*(P \circ \hat{P})+ W_2*U$, where $\hat{P}$ represents the estimate of $P$ based on $P'$ and $Q$ using $F$.

In order to solve the above \emph{optimization problem}, we need to find the \emph{optimal} anonymization $P'$ in the \emph{solution space} containing all possible anonymizations $P's$ that satisfy the fusion-resilient-anonymization property defined earlier. One way to look at this \emph{solution space} is to consider the set of all anonymizations possible by anonymizing $P$ to different \emph{levels}. Note that the definition of \emph{Anonymization Level} depends on the specific anonymization scheme to be employed. For example, in $K$-anonymization, the value of $k$ represents the anonymization level. The more the value of $k$ is, the more the anonymization level. As mentioned in Section 1, in our work, we use $K$-anonymization as the basic anonymization scheme. For a given dataset $P$, let $i$ denote the anonymization level and $P'_i$ denote the release obtained by anonymizing $P$ to level $i$. We use the \emph{discernibility metric} defined in~\cite{Bayardo2005} to measure the utility of a $k$-anonymized data set. The metric can be mathematically stated as follows.
\[C_{DM}(g,k)=\sum_{\forall|E|\geq k}^{}{|E|^{2}} + \sum_{\forall |E| < k}^{}{|D|*|E|}\]

where $E$ refers to the \emph{clusters} or \emph{equivalence classes} of the data set induced by $k$-anonymization of $g$ using the value $k$. The reader is referred to the original paper for further details. Based on the above definition, let the utility of $P'_i$ be denoted by $U_i$. The optimization function $H$ can now be defined based on anonymization level $i$ as: 
\[ H_i = W_1*(P \circ \hat{P_i})+ W_2*U_i \]

Let $T_u$ be the minimum utility required for the release. Now, the above generic problem statement can be instantiated as:

\textbf{Problem Statement:} 
Find $P'_{i_{opt}}$, such that
\[ H_{i_{opt}} = \max_{\forall{i}}{H_i}\]
where, $(P \circ \hat{P_i}) \geq T_p$ and $U_i \geq T_u$.

\section{Solution}

\begin{figure*}[!htp]
\centering
\epsfig{figure=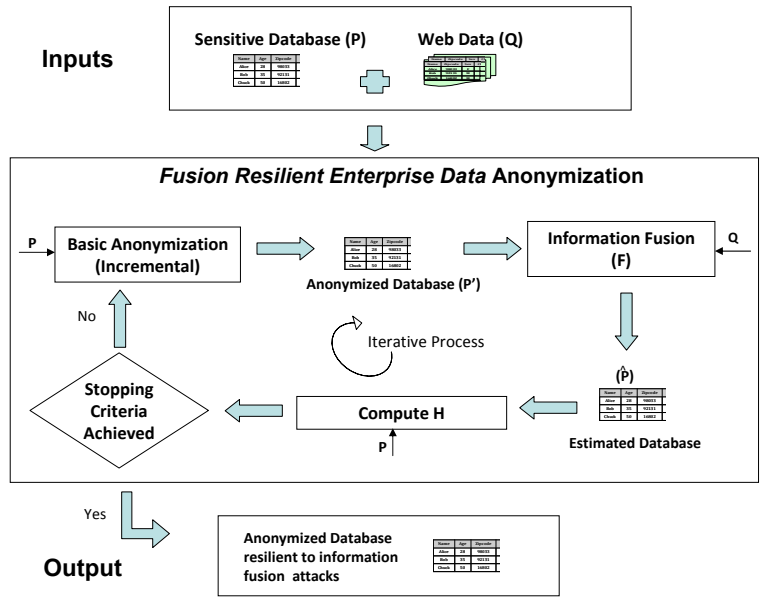, scale=0.5}
\caption{Fusion Resilient Enterprise Data Anonymization}
\label{blockdiagram}
\end{figure*}

In this section we propose a simple iterative algorithm to find the \emph{fusion resilient anonymization} for a given sensitive dataset. The strategy is to take any basic anonymization scheme such as $k$-anonymization and \emph{incrementally} anonymize the data. The level of anonymization is increased in steps (increase $k$ in steps), until the utility of the release falls below a threshold. In each step, the web-based fusion attack is simulated to find whether the resulting candidate anonymization offers enough protection. If yes, the candidate anonymization is retained, otherwise it is discarded. This results in a set of all \emph{candidate} anonymizations present in the \emph{solution space}. We then search for the optimal anonymization level that offers the maximum weighted sum of protection and utility. Figure~\ref{blockdiagram} illustrates our approach. 

Algorithm 1 presents this solution in procedural format as FRED\_Anonymization (Fusion Resilient Enterprise Data Anonymization). The algorithm uses the Basic\_Anonymization procedure that takes a sensitive data and level of anonymization as inputs and produces an anonymization of the input data to the corresponding anonymization level. For this, any basic anonymization algorithm such as the ones proposed in~\cite{Ferrer2002}~\cite{LeFevreDR06} can be used to generate a $k$-anonymization. Note that in case of $k$-anonymization the minimal level of anonymization is achieved by using the value $k=2$. The algorithm uses the Basic\_Anonymization procedure to anonymize the sensitive data for increasing values of the anonymization level $(level)$. The stopping condition for this loop is achieved when the utility of anonymized result ($P'$) denoted by $U_{level}$ falls below the threshold $T_u$. In each iteration, the algorithm simulates an information fusion attack to produce the estimate an adversary could obtain ($\hat{P}_{level}$). The \emph{dissimilarity} between the estimated values $\hat{P}_{level}$ and the original values $P$ is computed using the procedure Dissimilarity\_Measure which takes two datasets as input and outputs the dissimilarity value as described in Section 4. At this point, the dissimilarity is compared against a threshold value $T_p$ to check if the anonymization offers enough protection against information fusion attacks. If yes, the weighted sum of dissimilarity and utility is computed and stored as $H(i)$. Finally, the algorithm searches for the anonymization level $i$ that offers the maximum value for the weighted sum of protection and utility $H_{max}$. The anonymization $P'_{i_{opt}}$ corresponding to $H_{max}$ is the fusion resilient anonymization of the original data that offers maximum weighted protection as well as utility.

\begin{algorithm}
\caption{FRED\_Anonymization}
\label{algo1}
\begin{algorithmic}[1]
\STATE $P \leftarrow$ Sensitive Data
\STATE $Q \leftarrow$ Web Data 
\STATE $F \leftarrow$ Information Fusion System 
\STATE $T_p \leftarrow$ Protection Threshold 
\STATE $T_u \leftarrow$ Utility Threshold 
\STATE $W_1 \leftarrow$ Protection Weight 
\STATE $W_2 \leftarrow$ Utility Weight

\STATE $level \leftarrow -1$
\STATE $i \leftarrow 0$

\REPEAT
\STATE $level \leftarrow level + 1$
\STATE $P'_{level} \leftarrow$ Basic\_Anonymization($P$, level)
\STATE $\hat{P}_{level} \leftarrow F(P'_{level}, Q)$
\STATE $P \circ \hat{P}_{level} \leftarrow$ Dissimilarity\_Measure$(P, \hat{P}_{level})$
\STATE $U_{level} \leftarrow$ Utility($P'_{level}$)
\IF{$(P \circ \hat{P}_i) \geq T_p$}
\STATE $H(i) \leftarrow W_1*(P \circ \hat{P}_{level})+ W_2*U_{level}$
\STATE $i \leftarrow i+1$
\ENDIF
\UNTIL{$U_{level} \geq T_u$}

\STATE $i_{max} \leftarrow i-1$
\STATE $H_{max} \leftarrow H(0)$
\FOR{$i=1$ to $i=i_{max}$}
\IF{$H(i) \geq H_{max}$}
\STATE $i_{opt} \leftarrow  i$
\ENDIF
\ENDFOR

\RETURN $P'_{i_{opt}}$
\end{algorithmic}
\end{algorithm}
 
%

\section{Experimental Results}

In this section we present experimental results by demonstrating the \emph{web-based information-fusion attack} on a real-life dataset. The goals here are to quantify the information gained by the adversary through information fusion and demonstrate the FRED\_Anonymization algorithm.

\subsection{Setup}
The sensitive data ($P$) is collected from a real-life enterprise (a public university) and contains salary information and performance review numbers of the employees (faculty). The employee \emph{Salary} is the \emph{sensitive} attribute while the performance review numbers are the \emph{non-sensitive attributes}. The data is anonymized ($P'$) so as to \emph{suppress} all of the salary information and $k$-anonymize the non-sensitive attributes using \emph{microaggregation} based $k$-anonymization proposed in~\cite{Ferrer2002}. The external data($Q$) is collected from the employee web pages and external links from there. Based on domain knowledge, we formulate a simplistic set of knowledge rules to fuse $P'$ and $Q$ and build a fuzzy inference system to estimate the employee salary as illustrated in Figure ~\ref{newfis}. All the rules are assigned uniform weights. 

All the experiments were implemented using Matlab on a PC with Intel Pentium 4 (1.8GHz) processor and 1GB of RAM running Microsoft Windows XP. 



\subsection{Information Gain}
\begin{figure}[htb]
\centering
\epsfig{figure=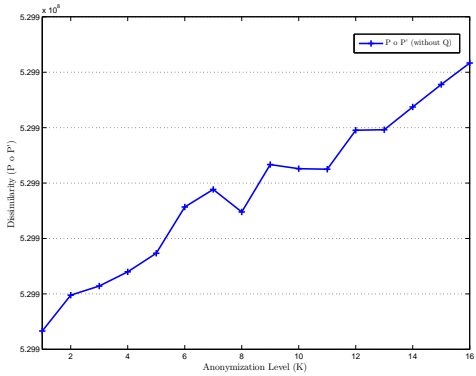, scale=0.3}
\caption{Before Information Fusion $(P\circ P')$}
\label{withoutq1}
\end{figure}

\begin{figure}[htb]
\centering
\epsfig{figure=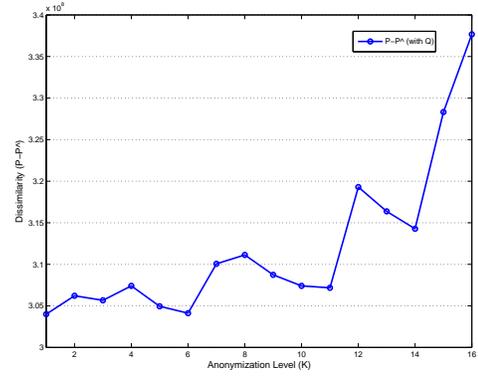, scale=0.3}
\caption{After Information Fusion $(P\circ\hat{P})$}
\label{withq}
\end{figure}

Our first study aims to quantify the \emph{information gain} obtained by the attacker in estimating the sensitive data $P$, by introducing web-based auxiliary information $Q$. Consider the adversary's knowledge of the original data at two stages 1. \emph{Before} information fusion, and 2. \emph{After} information fusion. Recall that to start with, the adversary has access to the anonymized release $P'$. The adversary then collects $Q$ and fuses this with $P'$ to obtain $\hat{P}$. So, before performing information fusion, the adversary's (best) knowledge about the original data is the anonymized version itself, i.e $P'$ (in the absence of $Q$). In this case, we have the dissimilarity between the original and the adversary's estimate $(P\circ \hat{P}) == (P\circ P')$. Figure ~\ref{withoutq1} plots this $(P\circ P')$ for increasing values of $k$. It is not surprising to observe that the \emph{dissimilarity} increases as $k$ increases, since the \emph{level} of anonymization increases with $k$. After performing information fusion, the adversary obtains $\hat{P}$ by fusing $P'$ with $Q$ using $F$. Figure ~\ref{withq} plots this $(P\circ\hat{P})$ for increasing values of $k$. Notice that $(P\circ\hat{P})$ is lesser than $(P\circ P')$ for all values of $k$. In other words, the estimate made by the attacker $(\hat{P})$ \emph{after} information fusion is closer to $(P)$ than when compared to the estimate available \emph{before} information fusion $(P')$. The difference between $(P\circ P')$ and $(P\circ\hat{P})$ is precisely the amount of information gained by the adversary through information fusion. Hence, the \emph{Information Gain} $G$ of the adversary is the difference between the closeness of the estimates available before and after information fusion. 
\[G = (P\circ P')- (P\circ\hat{P})\] 

\begin{figure}[htb]
\centering
\epsfig{figure=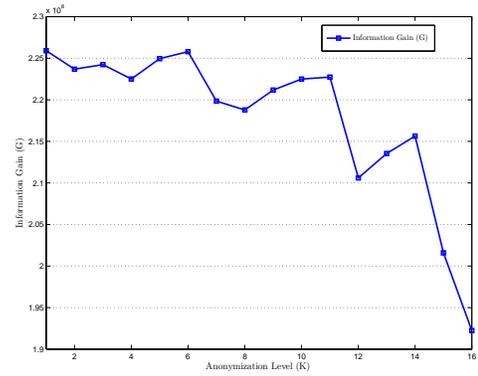, scale=0.3}
\caption{Information Gain $(G)$}
\label{infogain}
\end{figure}

Figure ~\ref{infogain} plots $G$ for increasing values of $k$. It is interesting to observe that $G$ does not necessarily increase with $k$. This implies that as the level of anonymization increases, the information gained by the attacker decreases. The reason for this is that as the level of anonymization increases, the input ($P'$) to the information fusion system gets worse and thus forces the system to output incrementally bad estimates. 


\subsection{Optimal Anonymization}
\begin{figure}[!htb]
\centering
\epsfig{figure=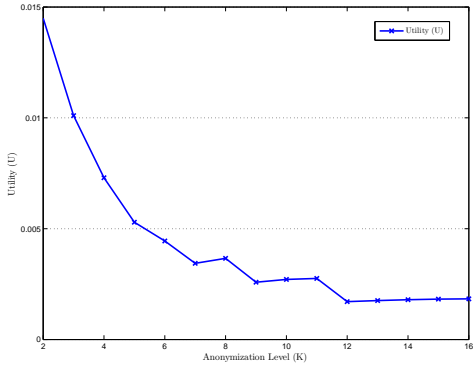, scale=0.3}
\caption{Utility $U_k$}
\label{utilvsk1}
\end{figure}

\begin{figure}[htb]
\centering
\epsfig{figure=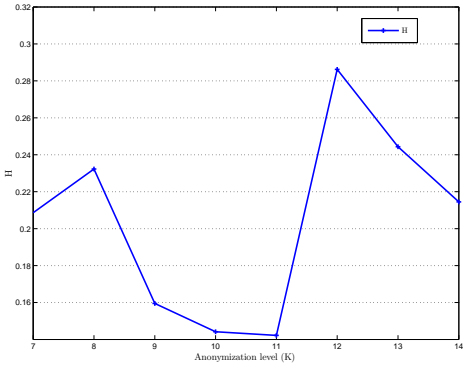, scale=0.3}
\caption{Weighted Sum Of Protection And Utility $H_k$}
\label{fvsk}
\end{figure}

We now study the fusion resilient enterprise data anonymization that leads to maximum weighted sum of protection and utility as formulated in Section 4. We use the \emph{discernibility metric} defined in ~\cite{Bayardo2005} to measure the utility of a $k$-anonymized data set. The basic idea here is to assign each data sample (or vector) a \emph{cost} based on the number of data vectors it is indistinguishable from, or in other words, the size of the cluster it falls into. If the cluster size it falls into is greater than $k$, then the cost assigned is equal to the size of the cluster. If the cluster size is less than $k$, then the cost is much severe (since it does not adhere to the definition of $k$-anonymity) and is equal to the product of the size of the whole data set and the size of the cluster. 
\[
C_i = \left\{
\begin{array}{lr}
|E|^{2} & if |E|\geq k \\
|D|*|E| & otherwise
\end{array}
\right\}
\]
Using this definition, we define the utility of the data set $U = \{u_{i1}\}_{m \times 1}$ as a column matrix where each entry is the inverse of the cost assigned to the corresponding data point. 
\[ u_{i1} = 1/C_i \]
To show how utility of the release varies with increasing level of anonymization (increasing values of $k$), we calculate the utility of the entire release using the discernibility definition ~\cite{Bayardo2005} as:
\[C_{DM}(k)=\sum_{\forall|E|\geq k}^{}{|E|^{2}} + \sum_{\forall |E| < k}^{}{|D|*|E|}\]
\[U_k = 1/C_{DM}(k)\]
Figure ~\ref{utilvsk1} plots $U_k$ for increasing values of $k$. It is straight-forward to observe that utility of data decreases as $k$ increases. The goal now is to find the optimal $k$ value such that the resulting anonymization offers maximum weighted sum of privacy protection and utility formulated as: 
\[H =  \frac{1}{m}*Tr((P\circ\hat{P})^T W_1 (P\circ\hat{P})) + \frac{1}{m}*Tr(U^T W_2 U)\]
We establish the threshold values for protection and utility as $T_p = 3.075$ $T_u = 0.0018$ based on experimental observations. For these threshold values, we obtain the solution space of $k = 7$ to $14$. We assign equal weights to privacy protection and utility i.e $W_1 = W_2 = 0.5$, $W_i = {0.5}_{m \times n}$i.e . Based on this setup, Figure ~\ref{fvsk} plots $H$ for increasing values of $k$ within the solution space. By running an optimization for the maximum value of $H$, we obtain the result $k=12$. This is the optimal $k$ value that provides the maximum weighted sum of protection and utility.


\section{Conclusion}
In this paper, we establish two problems encountered in privacy preserving enterprise data management:
\begin{enumerate}
\item Enterprise Data anonymization involves minimizing data disclosure in the presence of explicit individual-identifier information.
\item Existing anonymization techniques fall short in protecting enterprise data privacy in case of adversarial information fusion.
\end{enumerate}
We defined the \emph{Web-Based Information-Fusion Attack} where-in an adversary uses information fusion techniques to fuse anonymized data with publicly available information from the web to inflict a privacy breach. Our experimental demonstration of the attack present the practicality and easiness with which such attacks might lead to revelation of sensitive data. We formulate the problem of finding a \emph{fusion resilient data anonymization} and propose a simple solution to address this problem. While it is not possible to entirely prevent fusion based privacy attacks, one can minimize the extent of breach possible through intelligent data anonymization.

\bibliographystyle{IEEEtran}

\bibliography{icde1}

\end{document}